\documentclass[sigconf,screen]{acmart}

\usepackage{tabularx}
\usepackage{booktabs}
\usepackage{url}
\usepackage{xcolor}
\usepackage{multicol}
\usepackage{balance}
\usepackage[ruled,vlined,lined,commentsnumbered]{algorithm2e}
\usepackage{colortbl}
\usepackage[most]{tcolorbox}
\usepackage[normalem]{ulem}
\usepackage{listings}
\usepackage{comment}

\newboolean{showcomments}
 \setboolean{showcomments}{true}
\ifthenelse{\boolean{showcomments}}
    {\newcommand{\mynote}[2]{
      \fbox{\bfseries\sffamily\scriptsize#1}
        {\small$\blacktriangleright$\textsf{\emph{#2}}$\blacktriangleleft$}}}
        { \newcommand{\mynote}[2]{}}

\newcommand{\tool}{{\textsc{iTiger}}\xspace}
\newcommand{\tm}{{\texttt{Tampermonkey}}\xspace}

\setcopyright{acmlicensed}
\acmPrice{15.00}
\acmDOI{10.1145/3540250.3558934}
\acmYear{2022}
\copyrightyear{2022}
\acmSubmissionID{fse22demo-p92-p}
\acmISBN{978-1-4503-9413-0/22/11}
\acmConference[ESEC/FSE '22]{Proceedings of the 30th ACM Joint European Software Engineering Conference and Symposium on the Foundations of Software Engineering}{November 14--18, 2022}{Singapore, Singapore}
\acmBooktitle{Proceedings of the 30th ACM Joint European Software Engineering Conference and Symposium on the Foundations of Software Engineering (ESEC/FSE '22), November 14--18, 2022, Singapore, Singapore}

\begin{document}
\title{\textsc{iTiger}: An Automatic Issue Title Generation Tool}

\author{Ting Zhang}
\email{tingzhang.2019@phdcs.smu.edu.sg}
\authornote{Both authors contributed equally to this research.}
\affiliation{%
  \state{Singapore Management University}
  \country{Singapore}
}

\author{Ivana Clairine Irsan}
\authornotemark[1]
\email{ivanairsan@smu.edu.sg}
\affiliation{%
  \state{Singapore Management University}
  \country{Singapore}
}

\author{Ferdian Thung}
\email{ferdianthung@smu.edu.sg}
\affiliation{%
  \state{Singapore Management University}
  \country{Singapore}
}

\author{DongGyun Han}
\email{DongGyun.Han@rhul.ac.uk}
\affiliation{%
  \state{Royal Holloway, University of London}
  \country{United Kingdom}
}

\author{David Lo}
\email{davidlo@smu.edu.sg}
\affiliation{%
  \state{Singapore Management University}
  \country{Singapore}
}

\author{Lingxiao Jiang}
\email{lxjiang@smu.edu.sg}
\affiliation{%
  \state{Singapore Management University}
  \country{Singapore}
}


\renewcommand{\shortauthors}{Zhang et al.}

\begin{abstract}
In both commercial and open-source software, bug reports or issues are used to track bugs or feature requests.
However, the quality of issues can differ a lot.
Prior research has found that bug reports with good quality tend to gain more attention than the ones with poor quality.
As an essential component of an issue, title quality is an important aspect of issue quality.
Moreover, issues are usually presented in a list view, where only the issue title and some metadata are present.
In this case, a concise and accurate title is crucial for readers to grasp the general concept of the issue and facilitate the issue triaging.
Previous work formulated the issue title generation task as a one-sentence summarization task.
A sequence-to-sequence model was employed to solve this task.
However, it requires a large amount of domain-specific training data to attain good performance in issue title generation.
Recently, pre-trained models, which learned knowledge from large-scale general corpora, have shown much success in software engineering tasks.

In this work, we make the first attempt to fine-tune BART, which has been pre-trained using English corpora, to generate issue titles.
We implemented the fine-tuned BART as a web tool named \textsc{iTiger}, which can suggest an issue title based on the issue description.
\textsc{iTiger} is fine-tuned on 267,094 GitHub issues.
We compared \textsc{iTiger} with the state-of-the-art method, i.e., iTAPE, on 33,438 issues.
The automatic evaluation shows that \textsc{iTiger} outperforms iTAPE by 29.7\%, 50.8\%, and 34.1\%, in terms of ROUGE-1, ROUGE-2, ROUGE-L F1-scores.
The manual evaluation also demonstrates the titles generated by BART are preferred by evaluators over the titles generated by iTAPE in 72.7\% of cases.
Besides, the evaluators deem our tool as useful and easy-to-use. They are also interested to use our tool in the future.

\textbf{Demo URL:} \url{https://tinyurl.com/itiger-tool}

\textbf{Source code and replication package URL:} \url{https://github.com/soarsmu/iTiger}

\end{abstract}


\maketitle

\section{Introduction}
In software development and maintenance, bug reports or issues\footnote{Note in our work, we use the terms \emph{bug report} and \emph{issue} interchangeably.} are heavily used by developers to report bugs or propose new features.
With the proliferation of open-source software and social coding platforms, issue trackers are more accessible than ever.
On the one hand, given their varying experience levels, developers write bug reports with various qualities.
On the other hand, the quality of bug reports can impact the bug triaging process.
Prior research~\cite{guo2010characterizing} finds that well-written bug reports are more likely to gain triager’s attention and influence the decision on whether the bugs get fixed.

There is an emerging research interest in improving issue quality.
An issue usually includes a title and a description.
The description is optional and can contain extensive and rich information, such as detailed steps to reproduce a bug.
The title serves as the summary of the description.
Most prior works focus on improving the description of an issue and not its title~\cite{chaparro2017detecting}.
For instance, Chaparro et al.~\cite{chaparro2017detecting} consider that a good bug report description should clearly describe the Observed Behavior (OB), the Steps to Reproduce (S2R), and the Expected Behavior (EB). 
They propose an approach to improve bug description quality by alerting reporters about missing EB and S2R at reporting time.
Compared to the attention given to the quality of issue descriptions, the quality of issue titles has only recently received research interests.
Succinct and accurate issue titles can help readers quickly grasp the issue content and potentially speed up the bug triaging process, especially when issues are presented in the form of a list.
However, since developers may neglect to compose a succinct and accurate issue title, there is a need for an automatic issue title generation tool to help developers.

Chen et al.~\cite{chen2020stay} are the first to work on the issue title generation task, which aims to help developers write issue titles.
They formulated the issue title generation task as a one-sentence summarization task.
They propose iTAPE~\cite{chen2020stay}, which is a specialized tool to generate the issue title based on the issue description.
iTAPE relies on a sequence-to-sequence model~\cite{sutskever2014sequence} and is implemented with the OpenNMT framework\footnote{\url{https://github.com/OpenNMT/OpenNMT-py}}.
Unlike them, we leverage pre-trained models (PTMs), which have brought considerable breakthroughs in the field of artificial intelligence.
PTMs are pre-trained in much unlabeled data, and can be fine-tuned to solve downstream tasks.
Fine-tuning PTMs can usually achieve better performance than learning models from scratch~\cite{zhang2020sentiment,zt2022benchmark}.
To fill the gap of adopting PTMs to solve the issue title generation task, in this work, we leverage a type of PTMs, i.e., BART~\cite{lewis2019bart}, which has demonstrated promising performance in summarization and text generation tasks.
Specifically, we benefit from transfer learning, where the BART model we used has been pre-trained in large English corpora. The model is further fine-tuned using a GitHub issue title dataset collected by Chen et al.~\cite{chen2020stay}, which consists of 333,563 issues.

To build a bridge between research and practice, we present \textsc{iTiger}, a web-based tool to generate high-quality issue titles. To use \textsc{iTiger}, developers simply need to install a Userscript manager, i.e., \tm~\cite{Tampermo42:online}, that is available in most popular web browsers, such as Chrome, Safari, and Firefox.
After that, developers can install the script we provided.
\textsc{iTiger} is specially designed for the scenario when developers are creating a new issue: they can focus on drafting a detailed description, and \textsc{iTiger} can automatically generate a succinct and accurate summary of the description.
The script of \tool is published on GitHub.\footnote{\url{https://github.com/soarsmu/iTiger}}
\section{Approach}
\label{sec:approach}

\begin{figure*}[t]
\includegraphics[width=0.86\textwidth]{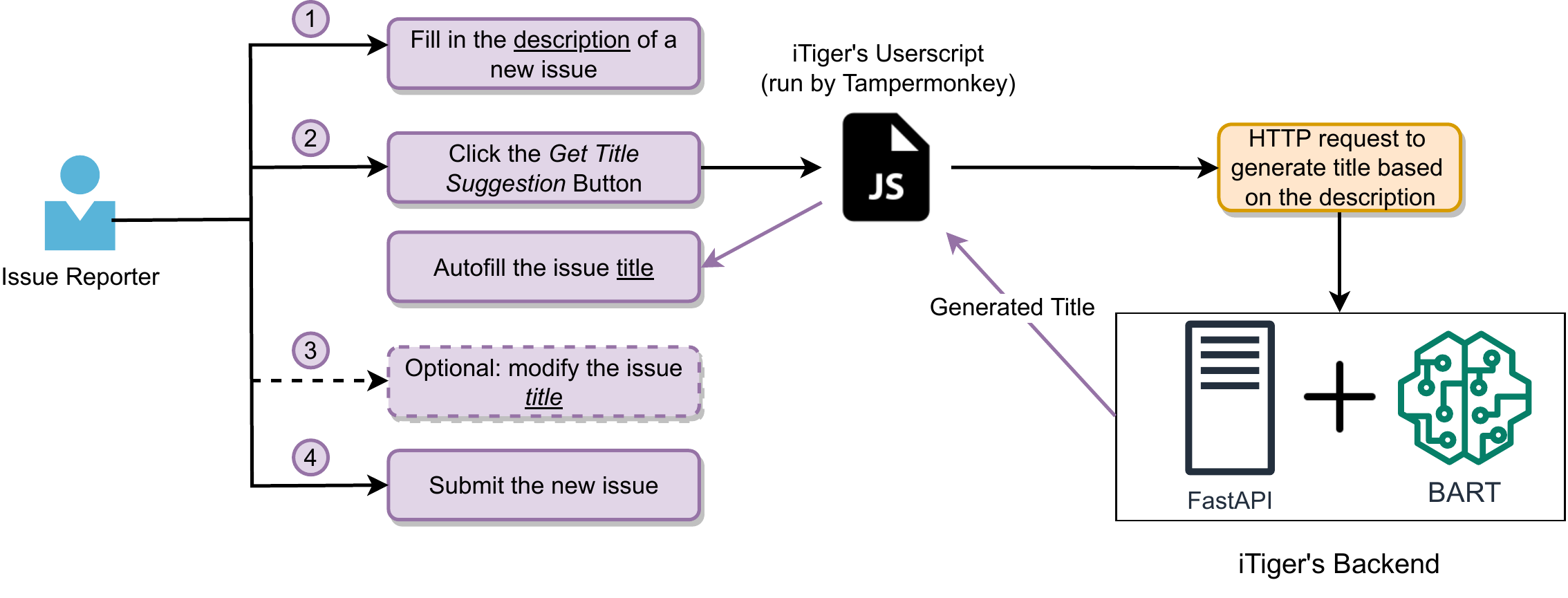}
\caption{The scenario of using \tool}
\label{fig:flow}
\end{figure*}

\begin{figure}[t]
\includegraphics[width=0.48\textwidth]{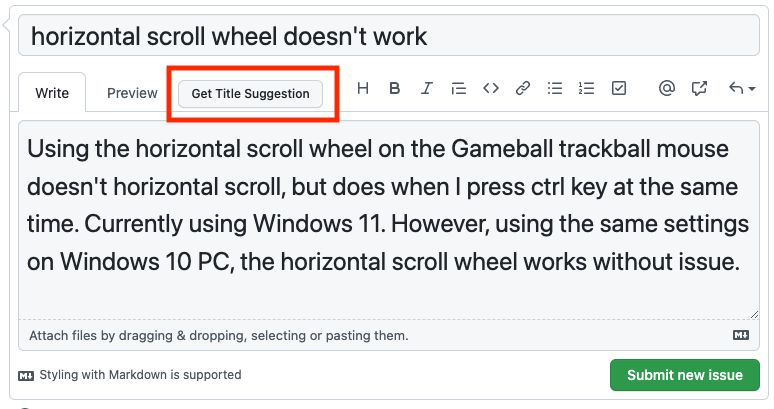}
\caption{The title field is filled in by \tool}
\label{fig:ui-after}
\end{figure}

We demonstrate the overall workflow of \tool in Figure~\ref{fig:flow}.
First, when issue reporters create a new issue, they need to fill in the description field of the issue (as shown in Figure~\ref{fig:ui-after}).
Second, upon finishing writing the issue description, they can click the \textit{Get Title Suggestion} button (as the red box shown in Figure~\ref{fig:ui-after}).
\tool then sends a request to the backend to generate the title.
After getting the title, \tool will auto-fill the title field of the issue.
Third, issue reporters can further modify and polish the issue title if they choose to do so.
Fourth, after they are satisfied with the issue title, they can proceed to open this new issue.

The underlying model of \tool is BART, a standard sequence-to-sequence (Seq2Seq) Transformer architecture that has been pre-trained by first corrupting a text with noising functions, and then learning to reconstruct the original text~\cite{lewis2019bart}.
Several noising functions are applied by BART, such as token masking, token deletion, and sentence permutation (sentences are shuffled in random order).
In the original paper, an evaluation on two news summarization datasets indicates that BART outperforms all existing works.
BART is pre-trained in the same corpora as RoBERTa, which includes over 160GB of uncompressed English text~\cite{liu2019roberta}.   
In this work, we adopt the base version of BART~\footnote{\url{https://huggingface.co/facebook/bart-base}} with 6 layers in the encoder and decoder.
BART was fine-tuned as a standard Seq2Seq model from the source sequence to the target sequence.
Specifically, in our task, the source sequence is the issue description, and the target sequence is the issue title.
We fine-tune BART by feeding it with the pairs of issue descriptions and titles.
In the inference stage, the input is the issue description and BART can generate the issue title as the output.
We leave all the hyper-parameters values to their defaults.
The detailed hyper-parameter settings are available in our replication package.
We trained BART on the issue description and title pairs extracted from GitHub.
The details about the data we used can be found in Section~\ref{sec:evaluation}.
Once the model is fine-tuned, we can utilize the model to generate the suggested title based on the issue description.
\section{Implementation Details}
\label{sec:implementation}

We have implemented \textsc{iTiger} as a Userscript which can be run in popular web browsers.
We believe, by seamlessly integrating \textsc{iTiger} with GitHub web UI (as opposed to making \textsc{iTiger} a standalone application), context switches between a standalone application and a browser are eliminated and therefore it saves developers time.
To generate issue titles more accurately, we first fine-tune the pre-trained BART on a domain-specific issue dataset.
We fine-tuned BART with 3 NVIDIA Tesla V100 GPUs.
Using fine-tuned BART model~\cite{lewis2019bart}, \textsc{iTiger} will generate a title suggestion based on the issue description.
Instead of directly submitting a new issue with the generated title, \textsc{iTiger} only fill the title field so the reporter has the freedom to modify the title if deemed necessary.

\textsc{iTiger}'s architecture consists of two components: frontend Userscript and backend. 
Frontend Userscript constructs the user interface. 
The backend provides a RESTful API for easy integration, such that a single deployment can serve multiple clients.

\textbf{Frontend Userscript:} The Userscript constructs \textsc{iTiger}'s user interface and integrates it with the backend service. The Userscript adds a \texttt{Get Title Suggestion} button on the new issue page of GitHub. When the button is clicked, the Userscript sends an HTTP request to the backend service and fills the title field with the suggested title from the backend service.

\textbf{Backend:} \textsc{iTiger}'s backend provides a RESTful API to serve the clients. It is developed using Python 3, utilizing the FastAPI framework. 
By leveraging server-client architecture, one backend service could serve multiple clients, and the clients do not need to store the model in their machines.
\textsc{iTiger}'s backend server passes the description to \textsc{iTiger}'s title generator model every time there is an incoming request.

\textbf{Deployment:} \textsc{iTiger}'s backend service is made available in our replication package.
It can be deployed in any machine via docker containerization by following the step-by-step guide provided in our replication package. 
Note that one backend service can serve multiple clients. 
To be able to use \textsc{iTiger}, the client's side (i.e., issue reporter) needs to install \tm extension in their browser and import our Userscript on it.
It will add a \texttt{Get Title Suggestion} button. 
When this button is clicked, it will send a request and fill the title field on GitHub's new issue page.
\section{Evaluation}
\label{sec:evaluation}

\subsection{Evaluation Setup}
\textbf{Dataset:} We adopt the publicly available issue title generation dataset provided by Chen et al.~\cite{chen2020stay}.
This dataset contains 333,563 issues collected from the Top-200 most-starred repositories on GitHub.
These issues have high-quality titles.
Three heuristic rules were adopted to select the issues: (1) issues whose titles have less than 5 words, or more than 15 words; (2) issues whose titles have more than 70\% words missing in the description; (3) issues whose titles have a sub-sequence can exactly match a particular part of the issue description, and this sub-sequence is over 70\% of the title length are filtered out.
More information about the dataset can be found in the previous work~\cite{chen2020stay}.
The training, validation, and test data were split by a ratio of 8:1:1.

\textbf{Automatic Evaluation:} We apply ROUGE metric~\cite{lin2004rouge} to evaluate the accuracy of generated issue titles.
Specifically, we report ROUGE-N (N=1,2) and ROUGE-L, which have been widely used in prior summarization papers~\cite{liu2019automatic,chen2020stay}. The recall, precision, and F1-score for ROUGE-N are calculated as follows:

\begin{equation*}
    Recall_{rouge-n} = \frac{count(overlapped\_N\_grams)}{count(N\_grams \in \mathit{ref\_summary})}
\end{equation*}
\begin{equation*}
    Precision_{rouge-n} = \frac{count(overlapped\_N\_grams)}{count(N\_grams \in \mathit{gen\_summary})}
\end{equation*}
\begin{equation*}
    F1_{rouge-n} = 2\times \frac{Recall_{rouge-n}\times Precision_{rouge-n}}{Recall_{rouge-n}+ Precision_{rouge-n}}
\end{equation*}

In the above equations, reference summary ($\mathit{ref\_summary}$) refers to the original issue title, while generated summary ($\mathit{gen\_summary}$) refers to the title generated by models.
ROUGE-1 and ROUGE-2 differ in whether we count uni-gram or bi-grams.
$\mathit{overlapped\_N\_grams}$ indicates the N-grams exist in both the reference summary and the generated summary.
Thus, using the above equations, Recall measures the percentage of the N-grams in the reference summary that has been covered by the generated summary, while Precision measures the percentage of N-grams in the generated summary that really exist in the reference summary.
F1-score considers both Precision and Recall.
Thus, in our work, we treat F1-score as the evaluation metric.
Slightly different from ROUGE-N, ROUGE-L F1-score is based on Longest Common Subsequence (LCS). 
It compares the similarity between two given texts in automatic summarization evaluation.
We report ROUGE-N (N=1,2), and ROUGE-L F1-score in our work and refer to them as ROUGE-1, ROUGE-2, and ROUGE-L.
To understand whether \tool can generate better titles than the state-of-the-art approach iTAPE can do, we report the results of the automatic evaluation on \tool and iTAPE.

\textbf{Manual Evaluation:} Since the goal of \tool is to help practical use, we conducted two types of manual evaluation: one focuses on comparing the accuracy between the title generated by BART and iTAPE, while the other focuses on the usability of the tool.

\noindent\textit{Accuracy:} We randomly sampled 30 issues from the test set.
We invited 5 evaluators: 3 Ph.D. students in Computer Science and 2 Research Engineers in Software Engineering.
They have a programming experience of more than 5 years and have been using GitHub for more than two years.
We provided 30 issue descriptions with two generated titles (one was generated by \tool, and the other was generated by iTAPE).
We also randomize the order of the titles produced by the two tools presented to the evaluators.
The evaluators do not know the authorship of the two titles.

\noindent\textit{Usability:} We asked the same 5 evaluators to install and use \tool directly on GitHub.
We asked them to provide three scores, each score ranging from 1 to 5 (strongly disagree, disagree, slightly agree, agree, and strongly agree).
The three scores evaluate three aspects: \tool is easy-to-use, \tool is useful, and they would like to use \tool in the future.

\subsection{Result}
\textit{Automatic Evaluation:} Table~\ref{tab:res} shows the ROUGE-1, ROUGE-2, and ROUGE-L F1-scores produced by the two models, including \tool and iTAPE.
Since we used the exact splits as iTAPE, we directly cite the results of iTAPE from its paper.
We observe that \tool outperforms iTAPE by 29.7\%, 50.7\%, and 34.1\%, in terms of ROUGE-1 F1 score, ROUGE-2 F1 score, and ROUGE-L F1 score, respectively.

\begin{table}[t]
    \caption {Results on automatic evaluation} 
    \label{tab:res} 
    \centering
    \begin{tabular}{@{}|l|lll|@{}}
        \hline
        \textbf{Approach} 
        & \textbf{ROUGE-1}
        & \textbf{ROUGE-2}
        & \textbf{ROUGE-L} \\
        \hline
        \textbf{iTAPE} & 31.36 & 13.12 & 27.79 \\
        \hline
        \textbf{\tool} & 40.67 & 20.6 & 37.26 \\
        \hline
    \end{tabular}
\end{table}

\begin{table}[t]
    \caption {Results on manual evaluation} 
    \label{tab:res-h1} 
    \centering
    \begin{tabular}{@{}|l|l|@{}}
        \hline
        \textbf{Approach} 
        & \textbf{\#Preferred} \\
        \hline
        \textbf{iTAPE} & 41 \\
        \hline
        \textbf{\tool} & 109 \\
        \hline
    \end{tabular}
\end{table}

\textit{Manual Evaluation:}
Table~\ref{tab:res-h1} shows the number of titles generated by the two tools that are preferred by evaluators.
The titles generated by \tool are preferred by evaluators on about 72.7\% cases.
In terms of usability, the evaluators consider \tool to be easy-to-use (5 out of 5) and useful (3.8 out of 5)
and are willing to use \tool in the future (4.6 out of 5).

\subsection{Threats to Validity}
Following prior works on summarization studies~\cite{zt2022prtiger,liu2019automatic}, we adopt both automatic evaluation (i.e., ROUGE metrics) and manual evaluation.
Like other manual evaluations, our experimental results may be biased. 
To minimize the potential biases, we invited 5 evaluators from our research group.
They have a programming experience of more than 5 years and have been using GitHub for more than two years.
The authorships of the issue titles were hidden when they conducted the evaluation.
\section{Related Work}
\label{sec:background}

\textbf{Issue Quality Understanding and Improvement.}
Existing research suggests that high-quality issues tend to get more attention than those with poor-quality, and easier-to-read bug reports have shorter lifetimes~\cite{bettenburg2008makes}.
Bettenburg et al.~\cite{bettenburg2008makes} conduct a survey among developers to investigate the quality of bug reports from developers' perspective.
They consider several fine-grained information in the issue description, such as code samples and stack traces.
Their findings include but not limited to the fact that bug reports containing stack traces get fixed sooner and easy-to-read reports are fixed faster.
Guo et al.~\cite{guo2010characterizing} also confirm the finding that high-quality bug reports are more likely to gain the triager’s attention, especially if they contain clear steps for reproducing the bug.

Although our task focuses only on a part of issues, i.e., generating issue title generation, our final goal is to help improve issue quality.
Our strategy is to suggest a high-quality issue title instead of detecting existing bad ones.
Our work complements the existing works that aim to improve bug report quality.

\vspace{0.1cm}\noindent\textbf{Software Artifact Generation.}
Our work on issue title generation belongs to a broader research topic of software artifact generation.
There have been several tools proposed to automatically generate different software artifacts, such as bug reports~\cite{rastkar2010summarizing,liu2020bugsum}, pull request titles~\cite{zt2022prtiger}, and pull request descriptions~\cite{liu2019automatic}.

Bug report summarization has attracted research interests for more than one decade~\cite{rastkar2010summarizing}.
Although it shares similarities with our task, there are some important differences.
An obvious difference is the length of the target sequence: bug report summaries usually contain several sentences, whereas an issue title is a single-sentence summary.
Recently, Liu et al.~\cite{liu2020bugsum} propose an unsupervised approach that first converts sentences to vectors and then leverages an auto-encoder network to extract semantic features.
They also utilize interactive discussions, i.e., comments of a bug report, to measure whether a sentence is approved or disapproved.

Furthermore, we recently worked on the task of pull request title generation~\cite{zt2022prtiger}.
We evaluate several state-of-the-art summarization approaches in the dataset we built.
Both the automatic and manual evaluation results indicate that the fine-tuned \texttt{BART-base} model achieved the best performance in the pull request title generation task.
We also provided an associated tool named \textsc{AutoPRTitle}~\cite{autopr2022}, which aims to automatically generate pull request titles.
Inspired by our earlier work~\cite{zt2022prtiger}, we apply pre-trained BART to the issue title generation task. 
\textsc{iTiger} differs from \textsc{AutoPRTitle} in several aspects: (1) the data source: Pull request title generation requires more data sources, including a pull request description, commit messages, and the associated issue titles.
In comparison, issue title generation only requires an issue description.
(2) the tool usage: We implement these two tools considering the different usages between them.
Since issue generation only requires an issue description, we implemented \textsc{iTiger} as a Userscript that embeds a user interface directly on the GitHub's issue creation page to reduce the context-switching between applications. 
On the other hand, pull request title generation require more data sources. Hence, we implemented \textsc{AutoPRTitle} as a stand-alone web application to better capture these different sources of information.
\section{Conclusion and Future Work}
\label{sec:conclusion}
In this paper, we present \textsc{iTiger}, which generates the issue title based on the issue description.
\textsc{iTiger} allows developers to modify the generated titles.
\textsc{iTiger} utilizes the state-of-the-art summarization model, i.e., BART.
In automatic evaluation, \textsc{iTiger} outperforms the prior specialized issue title generation model, i.e., iTAPE.
In manual evaluation, the evaluators indicate that they prefer the titles generated by BART over the ones generated by iTAPE in 73.1\% of cases.
They agree that \tool is easy to use and useful, and they are also willing to use our tool in their daily work.
In the future, we would like to train \textsc{iTiger} on larger and more data to improve the quality of generated issue titles.

\textbf{Acknowledgment}
This research / project is supported by the National Research Foundation, Singapore, under its Industry Alignment Fund – Pre-positioning (IAF-PP) Funding Initiative. Any opinions, findings and conclusions or recommendations expressed in this material are those of the author(s) and do not reflect the views of National Research Foundation, Singapore.

\balance

\bibliographystyle{ACM-Reference-Format}
\bibliography{main}


\begin{thebibliography}{16}


\ifx \showCODEN    \undefined \def \showCODEN     #1{\unskip}     \fi
\ifx \showDOI      \undefined \def \showDOI       #1{#1}\fi
\ifx \showISBNx    \undefined \def \showISBNx     #1{\unskip}     \fi
\ifx \showISBNxiii \undefined \def \showISBNxiii  #1{\unskip}     \fi
\ifx \showISSN     \undefined \def \showISSN      #1{\unskip}     \fi
\ifx \showLCCN     \undefined \def \showLCCN      #1{\unskip}     \fi
\ifx \shownote     \undefined \def \shownote      #1{#1}          \fi
\ifx \showarticletitle \undefined \def \showarticletitle #1{#1}   \fi
\ifx \showURL      \undefined \def \showURL       {\relax}        \fi
\providecommand\bibfield[2]{#2}
\providecommand\bibinfo[2]{#2}
\providecommand\natexlab[1]{#1}
\providecommand\showeprint[2][]{arXiv:#2}

\bibitem[\protect\citeauthoryear{??}{Tam}{[n.d.]}]%
        {Tampermo42:online}
 \bibinfo{year}{[n.d.]}\natexlab{}.
\newblock \bibinfo{title}{Tampermonkey Home Page}.
\newblock \bibinfo{howpublished}{\url{https://www.tampermonkey.net/}}.
\newblock
\newblock
\shownote{(Accessed on 05/21/2022).}


\bibitem[\protect\citeauthoryear{Bettenburg, Just, Schr{\"o}ter, Weiss,
  Premraj, and Zimmermann}{Bettenburg et~al\mbox{.}}{2008}]%
        {bettenburg2008makes}
\bibfield{author}{\bibinfo{person}{Nicolas Bettenburg}, \bibinfo{person}{Sascha
  Just}, \bibinfo{person}{Adrian Schr{\"o}ter}, \bibinfo{person}{Cathrin
  Weiss}, \bibinfo{person}{Rahul Premraj}, {and} \bibinfo{person}{Thomas
  Zimmermann}.} \bibinfo{year}{2008}\natexlab{}.
\newblock \showarticletitle{What makes a good bug report?}. In
  \bibinfo{booktitle}{\emph{Proceedings of the 16th ACM SIGSOFT International
  Symposium on Foundations of software engineering}}.
  \bibinfo{pages}{308--318}.
\newblock


\bibitem[\protect\citeauthoryear{Chaparro, Lu, Zampetti, Moreno, Di~Penta,
  Marcus, Bavota, and Ng}{Chaparro et~al\mbox{.}}{2017}]%
        {chaparro2017detecting}
\bibfield{author}{\bibinfo{person}{Oscar Chaparro}, \bibinfo{person}{Jing Lu},
  \bibinfo{person}{Fiorella Zampetti}, \bibinfo{person}{Laura Moreno},
  \bibinfo{person}{Massimiliano Di~Penta}, \bibinfo{person}{Andrian Marcus},
  \bibinfo{person}{Gabriele Bavota}, {and} \bibinfo{person}{Vincent Ng}.}
  \bibinfo{year}{2017}\natexlab{}.
\newblock \showarticletitle{Detecting missing information in bug descriptions}.
  In \bibinfo{booktitle}{\emph{Proceedings of the 2017 11th Joint Meeting on
  Foundations of Software Engineering}}. \bibinfo{pages}{396--407}.
\newblock


\bibitem[\protect\citeauthoryear{Chen, Xie, Yin, Ji, Chen, and Xu}{Chen
  et~al\mbox{.}}{2020}]%
        {chen2020stay}
\bibfield{author}{\bibinfo{person}{Songqiang Chen}, \bibinfo{person}{Xiaoyuan
  Xie}, \bibinfo{person}{Bangguo Yin}, \bibinfo{person}{Yuanxiang Ji},
  \bibinfo{person}{Lin Chen}, {and} \bibinfo{person}{Baowen Xu}.}
  \bibinfo{year}{2020}\natexlab{}.
\newblock \showarticletitle{Stay professional and efficient: Automatically
  generate titles for your bug reports}. In \bibinfo{booktitle}{\emph{2020 35th
  IEEE/ACM International Conference on Automated Software Engineering (ASE)}}.
  IEEE, \bibinfo{pages}{385--397}.
\newblock


\bibitem[\protect\citeauthoryear{Guo, Zimmermann, Nagappan, and Murphy}{Guo
  et~al\mbox{.}}{2010}]%
        {guo2010characterizing}
\bibfield{author}{\bibinfo{person}{Philip~J Guo}, \bibinfo{person}{Thomas
  Zimmermann}, \bibinfo{person}{Nachiappan Nagappan}, {and}
  \bibinfo{person}{Brendan Murphy}.} \bibinfo{year}{2010}\natexlab{}.
\newblock \showarticletitle{Characterizing and predicting which bugs get fixed:
  an empirical study of microsoft windows}. In
  \bibinfo{booktitle}{\emph{Proceedings of the 32Nd ACM/IEEE International
  Conference on Software Engineering-Volume 1}}. \bibinfo{pages}{495--504}.
\newblock


\bibitem[\protect\citeauthoryear{Irsan, Zhang, Thung, Lo, and Jiang}{Irsan
  et~al\mbox{.}}{2022}]%
        {autopr2022}
\bibfield{author}{\bibinfo{person}{Ivana~Clairine Irsan}, \bibinfo{person}{Ting
  Zhang}, \bibinfo{person}{Ferdian Thung}, \bibinfo{person}{David Lo}, {and}
  \bibinfo{person}{Lingxiao Jiang}.} \bibinfo{year}{2022}\natexlab{}.
\newblock \showarticletitle{AutoPRTitle: A Tool for Automatic Pull Request
  Title Generation}. In \bibinfo{booktitle}{\emph{2022 IEEE 38th International
  Conference on Software Maintenance and Evolution (ICSME)}}.
  \bibinfo{pages}{Tool Demo Track}.
\newblock


\bibitem[\protect\citeauthoryear{Lewis, Liu, Goyal, Ghazvininejad, Mohamed,
  Levy, Stoyanov, and Zettlemoyer}{Lewis et~al\mbox{.}}{2020}]%
        {lewis2019bart}
\bibfield{author}{\bibinfo{person}{Mike Lewis}, \bibinfo{person}{Yinhan Liu},
  \bibinfo{person}{Naman Goyal}, \bibinfo{person}{Marjan Ghazvininejad},
  \bibinfo{person}{Abdelrahman Mohamed}, \bibinfo{person}{Omer Levy},
  \bibinfo{person}{Veselin Stoyanov}, {and} \bibinfo{person}{Luke
  Zettlemoyer}.} \bibinfo{year}{2020}\natexlab{}.
\newblock \showarticletitle{{BART:} Denoising Sequence-to-Sequence Pre-training
  for Natural Language Generation, Translation, and Comprehension}. In
  \bibinfo{booktitle}{\emph{Proceedings of the 58th Annual Meeting of the
  Association for Computational Linguistics, {ACL} 2020, Online, July 5-10,
  2020}}. \bibinfo{publisher}{Association for Computational Linguistics},
  \bibinfo{pages}{7871--7880}.
\newblock


\bibitem[\protect\citeauthoryear{Lin}{Lin}{2004}]%
        {lin2004rouge}
\bibfield{author}{\bibinfo{person}{Chin-Yew Lin}.}
  \bibinfo{year}{2004}\natexlab{}.
\newblock \showarticletitle{Rouge: A package for automatic evaluation of
  summaries}. In \bibinfo{booktitle}{\emph{Text summarization branches out}}.
  \bibinfo{pages}{74--81}.
\newblock


\bibitem[\protect\citeauthoryear{Liu, Yu, Li, Guo, Wang, and Mao}{Liu
  et~al\mbox{.}}{2020}]%
        {liu2020bugsum}
\bibfield{author}{\bibinfo{person}{Haoran Liu}, \bibinfo{person}{Yue Yu},
  \bibinfo{person}{Shanshan Li}, \bibinfo{person}{Yong Guo},
  \bibinfo{person}{Deze Wang}, {and} \bibinfo{person}{Xiaoguang Mao}.}
  \bibinfo{year}{2020}\natexlab{}.
\newblock \showarticletitle{Bugsum: Deep context understanding for bug report
  summarization}. In \bibinfo{booktitle}{\emph{Proceedings of the 28th
  International Conference on Program Comprehension}}.
  \bibinfo{pages}{94--105}.
\newblock


\bibitem[\protect\citeauthoryear{Liu, Ott, Goyal, Du, Joshi, Chen, Levy, Lewis,
  Zettlemoyer, and Stoyanov}{Liu et~al\mbox{.}}{2019a}]%
        {liu2019roberta}
\bibfield{author}{\bibinfo{person}{Yinhan Liu}, \bibinfo{person}{Myle Ott},
  \bibinfo{person}{Naman Goyal}, \bibinfo{person}{Jingfei Du},
  \bibinfo{person}{Mandar Joshi}, \bibinfo{person}{Danqi Chen},
  \bibinfo{person}{Omer Levy}, \bibinfo{person}{Mike Lewis},
  \bibinfo{person}{Luke Zettlemoyer}, {and} \bibinfo{person}{Veselin
  Stoyanov}.} \bibinfo{year}{2019}\natexlab{a}.
\newblock \showarticletitle{Roberta: A robustly optimized bert pretraining
  approach}.
\newblock \bibinfo{journal}{\emph{arXiv preprint arXiv:1907.11692}}
  (\bibinfo{year}{2019}).
\newblock


\bibitem[\protect\citeauthoryear{Liu, Xia, Treude, Lo, and Li}{Liu
  et~al\mbox{.}}{2019b}]%
        {liu2019automatic}
\bibfield{author}{\bibinfo{person}{Zhongxin Liu}, \bibinfo{person}{Xin Xia},
  \bibinfo{person}{Christoph Treude}, \bibinfo{person}{David Lo}, {and}
  \bibinfo{person}{Shanping Li}.} \bibinfo{year}{2019}\natexlab{b}.
\newblock \showarticletitle{Automatic generation of pull request descriptions}.
  In \bibinfo{booktitle}{\emph{2019 34th IEEE/ACM International Conference on
  Automated Software Engineering (ASE)}}. IEEE, \bibinfo{pages}{176--188}.
\newblock


\bibitem[\protect\citeauthoryear{Rastkar, Murphy, and Murray}{Rastkar
  et~al\mbox{.}}{2010}]%
        {rastkar2010summarizing}
\bibfield{author}{\bibinfo{person}{Sarah Rastkar}, \bibinfo{person}{Gail~C
  Murphy}, {and} \bibinfo{person}{Gabriel Murray}.}
  \bibinfo{year}{2010}\natexlab{}.
\newblock \showarticletitle{Summarizing software artifacts: a case study of bug
  reports}. In \bibinfo{booktitle}{\emph{2010 ACM/IEEE 32nd International
  Conference on Software Engineering}}, Vol.~\bibinfo{volume}{1}. IEEE,
  \bibinfo{pages}{505--514}.
\newblock


\bibitem[\protect\citeauthoryear{Sutskever, Vinyals, and Le}{Sutskever
  et~al\mbox{.}}{2014}]%
        {sutskever2014sequence}
\bibfield{author}{\bibinfo{person}{Ilya Sutskever}, \bibinfo{person}{Oriol
  Vinyals}, {and} \bibinfo{person}{Quoc~V Le}.}
  \bibinfo{year}{2014}\natexlab{}.
\newblock \showarticletitle{Sequence to sequence learning with neural
  networks}.
\newblock \bibinfo{journal}{\emph{Advances in neural information processing
  systems}}  \bibinfo{volume}{27} (\bibinfo{year}{2014}).
\newblock


\bibitem[\protect\citeauthoryear{Zhang, Chandrasekaran, Thung, and Lo}{Zhang
  et~al\mbox{.}}{2022a}]%
        {zt2022benchmark}
\bibfield{author}{\bibinfo{person}{Ting Zhang}, \bibinfo{person}{Divya~Prabha
  Chandrasekaran}, \bibinfo{person}{Ferdian Thung}, {and}
  \bibinfo{person}{David Lo}.} \bibinfo{year}{2022}\natexlab{a}.
\newblock \showarticletitle{Benchmarking Library Recognition in Tweets}. In
  \bibinfo{booktitle}{\emph{2022 IEEE/ACM 30th International Conference on
  Program Comprehension (ICPC)}}. \bibinfo{pages}{343--353}.
\newblock
\urldef\tempurl%
\url{https://doi.org/10.1145/3524610.3527916}
\showDOI{\tempurl}


\bibitem[\protect\citeauthoryear{Zhang, Irsan, Thung, Han, Lo, and Jiang}{Zhang
  et~al\mbox{.}}{2022b}]%
        {zt2022prtiger}
\bibfield{author}{\bibinfo{person}{Ting Zhang}, \bibinfo{person}{Ivana~Clairine
  Irsan}, \bibinfo{person}{Ferdian Thung}, \bibinfo{person}{Donggyun Han},
  \bibinfo{person}{David Lo}, {and} \bibinfo{person}{Lingxiao Jiang}.}
  \bibinfo{year}{2022}\natexlab{b}.
\newblock \showarticletitle{Automatic Pull Request Title Generation}. In
  \bibinfo{booktitle}{\emph{2022 IEEE 38th International Conference on Software
  Maintenance and Evolution (ICSME)}}. \bibinfo{pages}{Research Track}.
\newblock


\bibitem[\protect\citeauthoryear{Zhang, Xu, Thung, Haryono, Lo, and
  Jiang}{Zhang et~al\mbox{.}}{2020}]%
        {zhang2020sentiment}
\bibfield{author}{\bibinfo{person}{Ting Zhang}, \bibinfo{person}{Bowen Xu},
  \bibinfo{person}{Ferdian Thung}, \bibinfo{person}{Stefanus~Agus Haryono},
  \bibinfo{person}{David Lo}, {and} \bibinfo{person}{Lingxiao Jiang}.}
  \bibinfo{year}{2020}\natexlab{}.
\newblock \showarticletitle{Sentiment Analysis for Software Engineering: How
  Far Can Pre-trained Transformer Models Go?}. In
  \bibinfo{booktitle}{\emph{2020 IEEE International Conference on Software
  Maintenance and Evolution (ICSME)}}. IEEE, \bibinfo{pages}{70--80}.
\newblock


\end{thebibliography}

\end{document}